 \documentclass[times,twocolumn]{article}

\usepackage{graphics}
\usepackage{graphicx}
\usepackage[numbers]{natbib}
\usepackage{epstopdf}
\usepackage{lscape}
\usepackage{amssymb}

\begin{document}

\title{Handedness asymmetry of spiral galaxies with z$<$0.3 shows cosmic parity violation and a dipole axis}


\author{Lior Shamir \\ Lawrence Technological University \\21000 W Ten Mile Rd. \\Southfield, MI 48075\\Phone:248-204-3512\\Email: lshamir@mtu.edu}

\date{}

\maketitle

\begin{abstract}
A dataset of 126,501 spiral galaxies taken from Sloan Digital Sky Survey was used to analyze the large-scale galaxy handedness in different regions of the local universe. The analysis was automated by using a transformation of the galaxy images to their radial intensity plots, which allows automatic analysis of the galaxy spin and can therefore be used to analyze a large galaxy dataset. The results show that the local universe (z$<$0.3) is not isotropic in terms of galaxy spin, with probability $P<\sim5.8\cdot10^{-6}$ of such asymmetry to occur by chance. The handedness asymmetries exhibit an approximate cosine dependence, and the most likely dipole axis was found at RA=132$^o$, DEC=32$^o$ with 1$\sigma$ error range of 107$^o$ to 179$^o$ for the RA. The probability of such axis to occur by chance is $P<1.95\cdot10^{-5}$ . The amplitude of the handedness asymmetry reported in this paper is generally in agreement with Longo, but the statistical significance is improved by a factor of 40, and the direction of the axis disagrees somewhat.
\end{abstract}




%\end{keyword}

%\end{frontmatter}

% \linenumbers

\section{Introduction}
\label{introduction}

Assuming that the universe is isotropic, it is expected that in a sufficiently large sector of the universe the number of galaxies that rotate clockwise will be roughly equal to the number of galaxies that rotate counterclockwise. However, recent evidence suggests that the local universe does not follow that expected balance, and show that the ratio between clockwise and counterclockwise galaxies in some regions is significantly different than 1:1, introducing galaxy handedness asymmetry.

Longo \citep{Lon11} analyzed the sense of rotation of 15,158 spiral galaxies imaged by Sloan Digital Sky Survey (SDSS) and showed cosmic parity violation and a possible cosmic dipole axis. The analysis was based on a relatively small dataset of galaxies classified manually by four undergraduate students, so that the number of galaxies in each 30$^o$ Right Ascension (RA) slice ranged between 0 to 3512 galaxies. Redshift of the galaxies was smaller than 0.085. The experiment was done using a web-based user interface that mirrored half of the galaxy images, so that the analysis was not biased by certain preferences of the human readers who classified them. A dipole axis was detected with RA=217$^o$, DEC=32$^o$ with a probability of occurring by chance of 7.9$\cdot10^{-4}$ \citep{Lon11}.

Asymmetry of clockwise and counterclockwise galaxies was also observed in the Galaxy Zoo dataset \citep{Lin10}. However, since the classification was done manually by a static graphical user interface, the asymmetry could be attributed to certain preferences of the Galaxy Zoo participants who used the web-based system.

The sense of rotation of a galaxy is a gross metrics that is relatively easy to measure, and it is not affected by atmospheric effects or hardware inaccuracy. However, analyzing large datasets of spiral galaxies manually requires significant labor, and might be biased by human errors or reader preferences. In this study a computer analysis was used to determine the handedness of spiral SDSS galaxies, and the results were analyzed in the light of the Cosmological Principle and the cosmological isotropy assumption.

\section{Image dataset}
\label{dataset}

The data used in the experiment are galaxy images collected by SDSS \citep{Yor00}, and classified by Galaxy Zoo \citep{Lin08}. The first full Galaxy Zoo data release contained 891,552 galaxies with redshift, classified manually by Galaxy Zoo participants. The galaxies that were classified manually as spiral galaxies were selected, and the Ganalyzer method described in Section~\ref{method} was used to reject the edge-on galaxies to provide a dataset that contains only spiral galaxies that their handedness can be determined. That provided a dataset of 352,705 spiral galaxies with identifiable sense of rotation. It should be noted that the galaxy population is not uniformly distributed in the night sky covered by SDSS, and some RA ranges have little or no galaxies.

The galaxy dataset of {\it Galaxy Zoo} was determined based on the human perception, and using a static user interface. Therefore, while the results of {\it Galaxy Zoo} provide a ``super-clean" galaxy dataset, they might be biased by human behavior and preferences. To avoid possible human bias, another dataset that was used in the experiment was the SpecObj view of SDSS DR7, which is based on the SpecObjAll table, but excludes bad and duplicate data. The galaxies in the SpecObj view were classified automatically using the Ganalyzer method \citep{Sha11}, providing a dataset of 345,599 spiral galaxies that are not edge-on. Declination of the galaxies was between $\sim$-11.2$^o$ to $\sim$70.3$^o$. The Galaxy Zoo dataset is available on-line through the Galaxy Zoo Data Release \citep{Lin10}, and the SpecObj view is available through SDSS CAS server.

\section{Image analysis method}
\label{method}

The handedness of each spiral galaxy was determined by transformation of the images to their radial intensity plots, as done by the Ganalyzer method \citep{Sha11,Sha11b}. To transform the galaxy image to its radial intensity plot, first the foreground pixels of the galaxy are separated from the background using the Otsu global threshold \citep{Ots79}. Then, the center of the galaxy is detected, and the radius is determined by the maximal distance between the object center and any foreground pixel \citep{Sha11}.

The radial intensity plot is a 360$\times$35 image, such that the value of the pixel $(x,y)$ is the median value of the 5$\times$5 windows around the pixel at image coordinates $(O_x+sin(\theta) \cdot r,O_y-cos(\theta)\cdot r)$ in the galaxy image, where $\theta$ is the polar angle (in degrees) and {\it r} is a radial distance. Intuitively, the radial intensity plot is an image of the radial intensities at different distances from the galaxy center. Each horizontal line in the radial intensity plot is then smoothed using a median filter with a span of 50 pixels. 

\begin{figure}[h]
%\figurenum{<text>}
%\epsscale{<num>}
\includegraphics[scale=0.46]{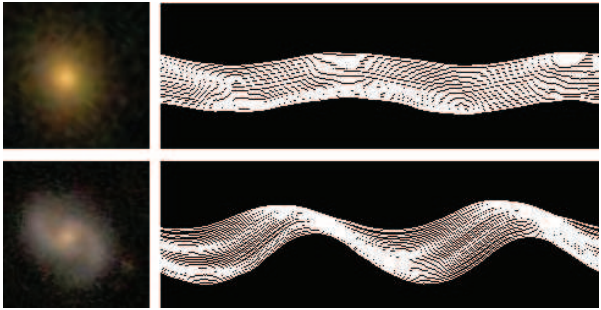}
%\plottwo{<epsfile>}{<epsfile>}
\caption{Galaxy images (left) and the transformation of the radial intensity plots such that the Y axis is the intensity and the X axis is the polar angle}
\label{radial}
\end{figure}

Figure~\ref{radial} shows the original galaxy image and a transformation of the radial intensity plot such that the Y axis is the intensity and the X axis is the polar angle. As the figure shows, in an image of a spiral galaxy the peaks are expected to shift due to the spirality of the arms, while in galaxies that are not spiral the peaks are expected to align in a near-straight line. The radial intensity plot transformation of galaxy images is described more thoroughly in \citep{Sha11}.

\begin{figure}[h]
%\figurenum{<text>}
%\epsscale{<num>}
\includegraphics[scale=0.64]{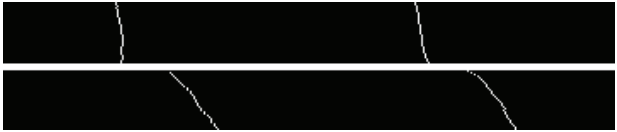}
%\plottwo{<epsfile>}{<epsfile>}
\caption{The peaks detected in the radial intensity plots of the elliptical galaxies of Figure~\ref{radial} }
\label{peaks}
\end{figure}

Once the radial intensity plot is generated, the peaks of each curve displayed in Figure~\ref{radial} are detected using a peak detection algorithm \citep{Mor00}, and the slope of the line created by the peaks can indicate whether the galaxy rotates clockwise or counterclockwise \citep{Sha11}. The slope is measured by first finding the peaks in the radial intensity curves displayed in Figure~\ref{radial}, and separating the peaks into groups such that each group contains peaks of the same arm of the spiral galaxy \citep{Sha11}. Figure~\ref{peaks} shows the peaks detected in the radial intensity plots of Figure~\ref{radial}.

For each group of peaks, the peaks are ordered by their Y coordinate in the radial intensity plot. Then, each peak at X coordinate {\it x} in the radial intensity plot is compared to the X coordinate of its next peak. If the X coordinate of the next peak is smaller than {\it x} then the counter {\it l} is incremented. If the X coordinate of the next peak is greater than {\it x}, the counter {\it r} is incremented. If after searching through all peaks $l>3r$, it is determined that the galaxy rotates clockwise. If $r>3l$ then the galaxy rotates counterclockwise. All other cases are considered undetermined handedness, and are excluded from the rest of the analysis. 

Using a set of 120 spiral galaxies taken from Galaxy Zoo and used in \citep{Sha09,Sha11}, and available for download at \newline http://vfacstaff.ltu.edu/lshamir/downloads/ganalyzer/GalaxyImages.zip, the method was tested for its accuracy and error rate. All galaxies had clear spirality, and the handedness, of each galaxy was classified manually by the author. The results showed that when using the criteria described above of $l>3r$ and $r>3l$ to determine the handedness, the handedness of 42.5\% of the galaxies was in agreement with the manual classification performed by the author, while the remaining 57.5\% of the galaxies had undetermined handedness. When using $l>2r$ and $r>2l$ as criteria the percentage of galaxies with determined handedness increased to $\sim$68.3\%, but with the sacrifice of $\sim$6.7\% error rate among those galaxies that their handedness was determined by the algorithm. When using $l>r$ and $r>l$ as criteria the determined handedness increased to 96\%, but the error rate also increased to 15\%.

Once the handedness of the galaxies is determined, the handedness asymmetry of a certain population of galaxies can be determined by $A=\frac{R-L}{R+L}$, where {\it A} is the asymmetry value, {\it L} is the number of galaxies rotating clockwise, and {\it R} is the number of galaxies rotating counterclockwise.

The accuracy of the Ganalyzer method is sensitive to redshift and magnitude when compared to human classification. However, Galaxy Zoo and the SpecObj view are not random samples, and contain bright objects that can be processed by Ganalyzer as described and analyzed in \citep{Sha11}. In lower redshift of $<$0.07 Ganayzer was able to determine the handedness ($l>3r$ or $r>3l$) of $\sim$42\% of the galaxies in the SpecObj view that were classified as spirals, while in the redshift ranges of 0.07-0.14 and 0.14-0.21 the detection rate was $\sim$38\% and $\sim$36\%, respectively. While the rate of the detection of the handedness is different at different redshift ranges, it is expected that in each redshift range the detection rate of clockwise rotating galaxies will be equal to counterclockwise rotating galaxies. The dependence on redshift can also be compared to manual classification of galaxies. While at redshift of 0.05 Ganalyzer classification into spiral and elliptical galaxies was in agreement with the manual classification in $\sim$89\% of the cases, at redshift 0.1 it was $\sim$87\%, and $\sim$82\% and $\sim$74\% at redshifts 0.15 and 0.25, respectively \citep{Sha11}. This, however, can be also attributed to the classification accuracy of the Galaxy Zoo citizen scientists, which drops as the redshift gets higher \citep{Lin10}. The redshift dependence of Ganalyzer accuracy \citep{Sha11,Doj13} can affect the magnitude of the signal of the asymmetry, but any inaccuracy is expected to be equal to both clockwise and counterclockwise directions, and therefore the direction of the asymmetry should not be affected. The Pearson correlation between the degree of spirality measured by Ganalyzer \citep{Sha11} and the apparent magnitude computed using $\sim$60,000 galaxies with redshift $<$0.3 provided a weak Pearson correlation of -0.036 \citep{Doj13}, with P$<10^{-8}$. That experiment showed that in the datasets of Galaxy Zoo and the SpecObj view, which contain brighter galaxies compared to the galaxies of SDSS DR7, there was merely a weak dependence between magnitude and the ability of Ganalyzer to detect spirality. Galaxies in which spirality was not detected were rejected and therefore did not have an effect on the analysis.

To test the response of Ganalyzer to changes in the brightness of the galaxies the experiment was repeated such that the $l>3r$ and $r>3l$ criteria were used, and the intensity of the foreground pixels of the galaxies was scaled down by different factors. As Table~\ref{intensity} shows, the percentage of galaxies with determined  handedness remains stable until the intensities of the foreground pixels are scaled down by more than 0.25. However, even when Ganalyzer has difficulties in determining the handedness, it does not missclassify left handed and right handed galaxies. The signal of the radial intensity plots weakens as the galaxy gets dimmer compared to its background, and therefore less peaks are detected when the galaxy is dimmer. However, the slope of the peaks determined by Ganalyzer \citep{Sha11} cannot be inverted because less peaks are found.

\begin{table}[h]
\centering
\begin{tabular}{|l|c|c|c|}
\hline
Intensity & Correct &  Incorrect & Undetermined \\
downscale & (\%) & (\%) & (\%) \\
\hline
1       &  42.5       & 0 & 57.5 \\    
0.75  &   42.5   &  0   & 57.5 \\ 
0.5     &  42.5  &  0  & 57.5   \\ 
0.25   &   42.5   &  0   & 57.5 \\ 
0.15  &   28.3   &   0    & 71.7 \\
0.1   &  17.5     &   0    & 82.5 \\
0.05  &   0        &   0    & 100 \\
\hline
\end{tabular}
\caption{Classification accuracy when the intensity of the foreground pixels of the galaxies is scaled down}
\label{intensity}
\end{table}

\section{Results}
\label{results}

The method described in Section~\ref{method} was applied to the two galaxy datasets described in Section~\ref{dataset}. The analysis was based on galaxies that met the handedness criteria described in Section~\ref{method}, so that Galaxy Zoo and the SpecObj view provided 126,501 and 121,414  spiral galaxies with determined directionality, respectively. 

\begin{figure}[h]
\includegraphics[scale=0.45]{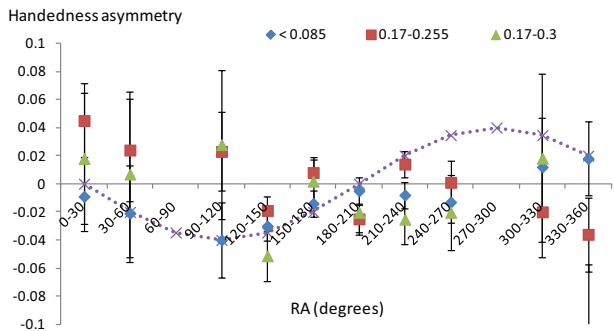}
\caption{Galaxy handedness asymmetry in different RA ranges and redshift ranges using SpecObj galaxies. The dotted line shows the cosine curve multiplied by -0.04 and starts at 132$^o$, to show cosine dependence as will be discussed later in the paper}
\label{by_redshift_3_ranges}
\end{figure}

Figure~\ref{by_redshift_3_ranges} shows the handedness asymmetry of the SpecObj galaxies defined in different RA ranges and three redshift ranges, which are $<$0.085 \citep{Lon11}, 0.085-0.17, and 0.17-0.3. The standard error was computed based on normal distribution using $1/\sqrt{N}$, where N is the total number of galaxies in the region and redshift range. The RA ranges [60,90] and [270,300] are empty due to the very low number of galaxies (70 and 0, respectively) in these RA slices.

As the figure shows, different RA and redshift ranges show different galaxy handedness asymmetries, suggesting that the local universe is not isotropic. Partial correlation can be observed between the redshift ranges in some of the RA regions, but the high standard errors in the different RA/redshift bins do not allow statistically significant conclusions about such correlation. Also, as described in Section~\ref{method} there is dependence between the redshift of the galaxies and the ability of Ganalyzer to detect their spirality and determine their handedness.

\begin{figure}[h]
\includegraphics[scale=0.40]{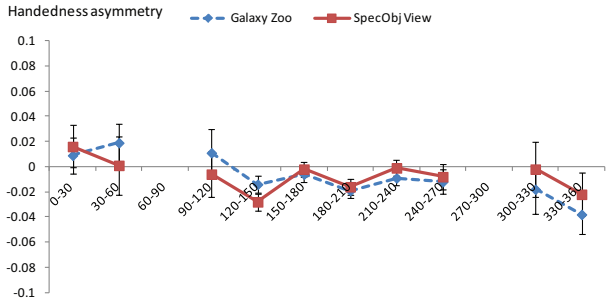}
\caption{Galaxy handedness asymmetry in different RA ranges}
\label{by_RA}
\end{figure}

Figure~\ref{by_RA} shows the galaxies handedness asymmetry defined in different regions using both Galaxy Zoo dataset and the galaxies from SpecObj.

The differences between Galaxy Zoo and SpecObj can be explained by the perceptional bias of the Galaxy Zoo dataset, in which the human-based classifications are biased towards left-handed spirals by $\sim$10\%. However, it is clear that the two datasets provide the same profile of galaxy handedness asymmetry. The strongest asymmetry of the SpecObj data was 0.028$\pm0.00683$, found in the RA range [120$^o$,150$^o$]. Assuming that the probability of a galaxy in that RA range to rotate clockwise is equal to its probability to rotate counterclockwise, given the standard error the probability to observe asymmetry of 0.028 by chance is $\sim4.3\cdot10^{-5}$. Considering just the four RA slices with more than 20,000 galaxies in them (the fifth most populated RA slice has just 9907), the probability that one or more of them have asymmetry equal or greater than 0.028 and one or more other slice have asymmetry of 0.16 (as observed in 180$^o$-210$^o$) or greater given the standard errors in these RA slices is $\sim5.8\cdot10^{-6}$.

As a test for bias, the analyses were repeated with the dataset described in Section~\ref{dataset} such that randomly selected half of the galaxies were mirrored. The handedness asymmetry values in the different RA ranges in the SpecObj and Galaxy Zoo data were smaller compared to the original data, and in all RA ranges the asymmetry was within the standard error, indicating that no systematic error was present. The largest handedness asymmetry measured with the mirrored data was 0.009$\pm0.01683$.

A real spiral spin preference would present itself as a dipole with a cos $\phi$  dependence, where $\phi$  is the space angle between the position of the galaxy and the preferred axis \citep{Lon11}. The RA, DEC and asymmetry values of each galaxy were used to find the most likely dipole axis using cosine dependence \citep{Lon11}. Each galaxy $i$ in the dataset was assigned with $\{-1,1\}$ handedness $d_i$, and the $\phi$ angle was computed for each galaxy as the difference between the angle of the galaxy and the angle of the candidate dipole axis. The $d\cdot|\cos \phi|$ was then fitted to $\cos \phi$ for all possible RA, DEC combinations in an exhaustive grid search such that the declination range was [-10$^o$,70$^o$], within the sky covered by SDSS, to find the most likely dipole axis with which the galaxy handedness asymmetries have cosine dependence.

\begin{figure}[h]
\includegraphics[scale=0.52]{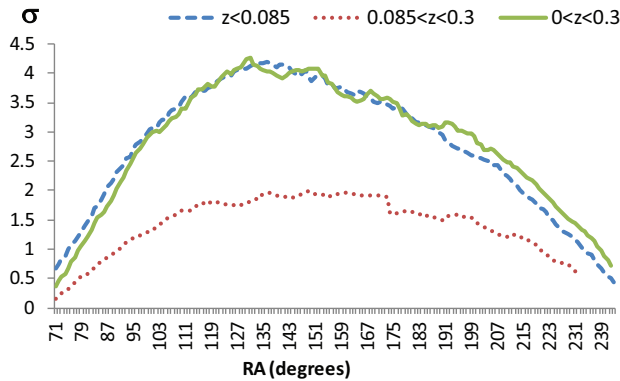}
\caption{$\sigma$ between the mean $\chi^2$ computed with 1000 attempts of fitting all galaxies with randomized handedness values to dipole axes with varying RA and declination 32$^o$}
\label{chi_square_ra}
\end{figure}

To deduce the most likely dipole axis, random handedness values were assigned to the galaxies, and the random handedness were fitted to $\cos \phi$ for each right ascension and declination axis, and repeated 1000 times for each axis to compute its mean and standard deviation of the $\chi^2$ computed with random galaxy handedness. Then, the $\chi^2$ computed for the same axis with the observed galaxy handedness determined by the Ganalyzer method as described in Section~\ref{method} was compared with the mean and $\sigma$ of the $\chi^2$ computed for that  axis using the 1000 runs of random galaxy handedness values. The axis that had the highest difference between the random and observed galaxy handedness in terms of $\sigma$ was found around RA=132$^o$, and DEC=32$^o$. Figure~\ref{chi_square_ra} shows the difference in terms of $\sigma$ between the observed $\chi^2$ and the mean $\chi^2$ of 1000 attempt of fitting randomized handedness values to dipole axes with varying RA and declination 32$^o$.

As the figure shows, the most likely dipole axis is at RA=132$^o$, and the 1$\sigma$ error range for the dipole is between RA 107$^o$ and 179$^o$. With $\sim4.27\sigma$ the probability to have such axis by chance is  $P<1.95\cdot10^{-5}$, showing that it is likely that galaxy handedness asymmetry form a dipole axis in the local universe ($z<0.3$). Handedness asymmetry in the 30$^o$ RA range around the most likely dipole [117$^o$,147$^o$] is $\sim$-0.017, and the asymmetry is $\sim$-0.027 in the same RA range where z$<$0.085. The declination of the most likely dipole axis is in agreement with the most likely dipole axis proposed by \citep{Lon11}, but the RA of 132$^o$ is different from the RA=217$^o$ \citep{Lon11}. However, the error range of the most likely dipole was $\pm$35$^o$ \citep{Lon11}, within the 1$\sigma$ statistical error range reported here.

%To better characterize the dipole, cosine dependence was tested in the same fashion for small redshift $<$0.085 and higher redshift range of 0.085-0.3. As Figure~\ref{chi_square_redshift} shows, 
As the graph also shows, cosine dependence was observed in low (z$<$0.085) and higher (0.085$<$z$<$0.3) redshift ranges, and the dipole axis is consistent in both ranges. The signal for the dipole is far stronger for z$<$0.085. The weaker signal at the higher redshifts can be attributed to the ability of Ganalyzer to accurately detect spirality, which decreases as the redshift of the galaxies gets higher \citep{Sha11,Doj13}. The most likely dipole axis in z$<$0.085 is at RA=136$^o$, with probability of $\sim7.19\cdot10^{-5}$ of having such cosine dependency by chance.

%\begin{figure}[h]
%\includegraphics[scale=0.58]{chi_square_z.eps}
%\caption{$\sigma$ between the mean of random and actual $\chi^2$ in different redshift ranges such that the declination is set to 32$^o$}
%\label{chi_square_redshift}
%\end{figure}

\begin{figure}[h]
%\figurenum{<text>}
%\epsscale{<num>}
\includegraphics[scale=0.52]{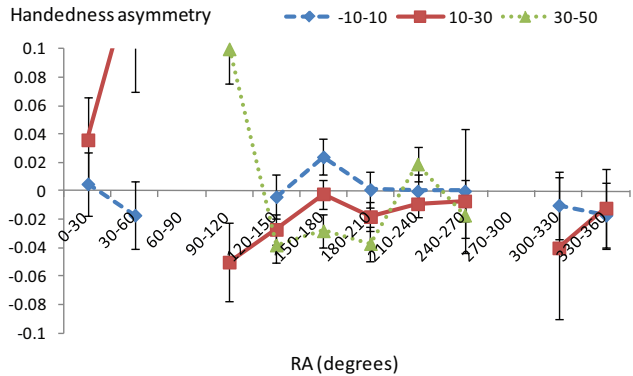}
%\plottwo{<epsfile>}{<epsfile>}
\caption{Galaxy handedness asymmetry in different RA and DEC ranges. DEC ranges are [-10$^o$,10$^o$], [10$^o$,30$^o$], [30$^o$,50$^o$]}
\label{by_declination}
\end{figure}

The handedness asymmetry was also analyzed for different declination ranges. The declination ranges that were used are  [-10$^o$,10$^o$], [10$^o$,30$^o$], and [30$^o$,50$^o$] of the SpecObj view galaxy dataset, which have 31,943, 46,482, and 33,006 galaxies in each declination range, respectively. The asymmetry values at different declination ranges and RA slices are displayed in Figure~\ref{by_declination}.

The limited declination ranges covered by SDSS do not allow comparison of declination ranges in opposite hemispheres, and not all declination and RA ranges of SDSS have galaxies in them. The highest number of galaxies in these declination ranges are in the RA range of [150$^o$,210$^o$], which contained 69,986 galaxies. As Figure~\ref{by_declination} shows, the galaxy handedness asymmetry in that RA range increases with the declination, while the asymmetry in other RA slices such as [210$^o$,240$^o$] the galaxies in the higher declination are biased toward counterclockwise rotation.

\section{Conclusion}
\label{conclusion}

The results of the experiment show that galaxy handedness asymmetry is different in different RA ranges within z$<$0.3, and the asymmetries exhibit a cosmological dipole axis. According to these observations, the local universe is not isotropic, meaning that the observed physical characteristics of the universe are different in different directions of observation. These results are in agreement with \citep{Lon11}. The most likely dipole axis was found at RA=132$^o$, DEC=32$^o$. This axis is somewhat different from the most likely dipole axis reported by \citep{Lon11} at RA=217$^o$, DEC=32$^o$ with an uncertainty of $\sim$35$^o$, but the difference is within the error ranges.

The limited range of declination in the SDSS dataset as well as the absence of data in certain right ascension ranges did not allow comparing all opposite hemispheres for handedness asymmetry for the purpose of fully profiling the characteristics of the handedness asymmetry in the local universe. It should be noted that the experiment is limited to the local universe, although the range of z$<$0.3 exceeds far beyond the scale of a galaxy supercluster. Higher-resolution analysis of the cosmic parity violation in the more distant universe will be possible when the Large Synoptic Sky Survey (LSST) starts operating, providing a much larger galaxy dataset and seeing deeper space.

The experiment is based solely on data acquired by Sloan Digital Sky Survey, and under the assumption that these raw data are not biased. A possible source of bias might be the automatic galaxy detection algorithm that is part of the SDSS pipeline, that might have a preference to galaxies of some certain directionality. This, however, conflicts with the observation that some of the opposite hemispheres have opposite asymmetries. In any case, even in the case that such bias exists, it is expected to be consistent for all SDSS data, so that the asymmetry values might be shifted, but the asymmetry profile itself should not change. Another possible weakness is that galaxies from different RA were taken at different times, and any change made to SDSS or its processing algorithms during that time could potentially lead to a biased dataset, and different galaxy asymmetries at different redshifts. It should also be noted that the SpecObj view and the Galaxy Zoo dataset contain bright objects, and are not random samples of the galaxies in DR7. A bias in the selection of these objects can also affect the results. 

Clearly, this study does not offer explanations neither to the nature of asymmetry at the cosmological scale, nor to the observation that the asymmetry changes at different regions. %The violation of the Cosmological Principle, however, might suggest that the different large structures of universe may not be in mechanical equilibrium, indicating that it might be possible that the universe is not an independent discrete structure, but part of a larger system.

\section{Acknowledgments}
I would like to thank John Wallin for his assistance in obtaining the redshift data for Galaxy Zoo galaxies, and the  reviewer for insightful comments that significantly improved the paper. Funding for the SDSS and SDSS-II has been provided by the Alfred P. Sloan Foundation, the Participating Institutions, the National Science Foundation, the US Department of Energy, the National Aeronautics and Space Administration, the Japanese Monbukagakusho, the Max Planck Society, and the Higher Education Funding Council for England. The SDSS Web Site is http://www.sdss.org/. The SDSS is managed by the Astrophysical Research Consortium for the Participating Institutions. The Participating Institutions are the American Museum of Natural History, Astrophysical Institute Potsdam, University of Basel, University of Cambridge, Case Western Reserve University, University of Chicago, Drexel University, Fermilab, the Institute for Advanced Study, the Japan Participation Group, Johns Hopkins University, the Joint Institute for Nuclear Astrophysics, the Kavli Institute for Particle Astrophysics and Cosmology, the Korean Scientist Group, the Chinese Academy of Sciences (LAMOST), Los Alamos National Laboratory, the Max Planck Institute for Astronomy (MPIA), the Max Planck Institute for Astrophysics (MPA), New Mexico State University, Ohio State University, University of Pittsburgh, University of Portsmouth, Princeton University, the United States Naval Observatory and the University of Washington.



%% References


% \bibliography{<your-bib-database>}

\begin{thebibliography}{00}


\bibitem
[Longo(2011]
{Lon11}
M. Longo, Phys. Lett. B, 699 (2011) 224--229.
%
\bibitem
[Lintott et al.(2010)]
{Lin10}
C. J. Lintott, et al., MNRAS, 410 (2010) 166.
%
\bibitem
[York et al.(2000)]
{Yor00}
D.G. York, et al., Astron. J., 120 (2000) 1579.
%
\bibitem
[Lintott et al.(2008)]
{Lin08}
C. J. Lintott, C. J., et al., MNRAS, 389 (2008) 1179.
%
\bibitem
[Shamir(2011a)]
{Sha11}
L. Shamir, Astrophys. J., 736 (2011a) 141.
%
\bibitem
[Shamir(2011b)]
{Sha11b}
L. Shamir, Astrophysics Source Code Library (2011b) 1105.011.
%
\bibitem
[Otsu(1979)]
{Ots79}
N. Otsu, IEEE Trans. Sys. Man. Cyber. 9 (1979) 62.
%
\bibitem[Morhac et al.(2000)]
{Mor00}
M. Morhac, et al., Methods in Research Physics A, 443 (2003) 108.
%
\bibitem
[Shamir(2009)]
{Sha09}
L. Shamir, MNRAS, 399 (2009) 1367.
%
\bibitem
[Dojcsak \& Shamir(2013)]
{Doj13}
L. Dojcsak, L. Shamir, ``Quantitative analysis of spirality in elliptical galaxies", submitted.
%
%http://arxiv.org/abs/hep-ph/0208208
%
\end{thebibliography}



\end{document}